# Improved assessment of the accuracy of record linkage via an extended *MaCSim* approach

Shovanur Haque and Kerrie Mengersen
Queensland University of Technology
shovanur.haque@hdr.qut.edu.au, k.mengersen@qut.edu.au

**Abstract**
Record linkage is the process of bringing together the same entity from overlapping data sources while removing duplicates. Huge amounts of data are now being collected by public or private organizations as well as by researchers and individuals. Linking and analysing relevant information from this massive data reservoir can provide new insights into society. However, this increase in the amount of data may also increase the likelihood of incorrectly linked records among databases. It has become increasingly important to have effective and efficient methods for linking data from different sources. Therefore, it becomes necessary to assess the ability of a linking method to achieve high accuracy or to compare between methods with respect to accuracy. In this paper, we improve on a Markov Chain based Monte Carlo simulation approach (*MaCSim*) for assessing a linking method. *MaCSim* utilizes two linked files that have been previously linked on similar types of data to create an agreement matrix and then simulates the matrix using a proposed algorithm developed to generate re-sampled versions of the agreement matrix. A defined linking method is used in each simulation to link the files and the accuracy of the linking method is assessed. The improvement proposed here involves calculation of a similarity weight for every linking variable value for each record pair, which allows partial agreement of the linking variable values. A threshold is calculated for every linking variable based on adjustable parameter 'tolerance' for that variable. To assess the accuracy of linking method, correctly linked proportions are investigated for each record. The extended *MaCSim* approach is illustrated using a synthetic dataset provided by the Australian Bureau of Statistics (ABS) based on realistic data settings. Test results show higher accuracy of the assessment of linkages.

# 1. Introduction

The advancement of modern record linkage methods came from three disciplines: statistics, computer science, and operations research. The basic ideas of record linkage were first introduced by the geneticist Howard Newcombe and his collaborators (Newcombe et al. 1959, 1962). In their work, they described odds ratios of frequencies and decision rules that identified matches and non-matches. Fellegi and Sunter (1969) provided a theoretical foundation of record linkage based on the ideas of Newcombe et al. (1959, 1962). They described how to estimate matching parameters directly from the files being matched. Their estimation methods consider three matching variables and assume conditional independence of agreement of linking variables. Their theory still remains the basis of many record linkage systems.

These early ideas have generated a wealth of theory, methods, algorithms and software systems for record linkage. Despite these advances, there are still challenges in linking records and it is still not easy to estimate matching parameters and error rates automatically, as highlighted by the work of Larsen and Rubin (1999). Importantly, current algorithms and software systems display varying levels of accuracy, but these are often not well communicated and are difficult to assess. This motivates the development of methods to assess the accuracy of record linking methods.

The most commonly used methods in record linkage are deterministic and probabilistic linkage methods. In a deterministic approach, two records are said to be a link if they agree on a high quality identifier (e.g. social security number, tax file number, driver license, etc.) or a combination of identifiers (e.g. first name, date of birth and street name). In probabilistic linking, no unique identifier is available, record pairs are compared with the values of linking variables that are common to both files. The individual variables used for connecting records are generally called linking variables or linking fields. In probabilistic linking method, each record pair is given a weight based on the likelihood that they are a match. For a record pair, the weight is determined by assessing every linking variable value for agreement or disagreement. Based on this assessment a weight is assigned for each linking variable and summing these individual weights over all linking variables for that pair. This summation is based on the premise of conditional independence, which means that for a record pair the agreement on a linking field is independent of agreement on any other linking field for that pair. Finally, a decision rule typically based on a cut-off value (or threshold) determines whether a record pair is linked or non-linked or should be considered further as a possible link (Fellegi and Sunter 1969). Probabilistic record linkage methods are now being well accepted and widely used (Sadinle 2013, 2014, 2016; Sayers et al. 2016, Steorts 2015, 2016; Herzog et al. 2007; Winkler 2001, 2005). In the current world scenario of data availability, probabilistic method is preferred for data linking purposes. In this paper, we used probabilistic linking approach.

Perfect linkage occurs when all matches are linked and all non-matches remain unlinked. In reality, it is very unlikely to have perfect linkage and linkage errors occur. Errors can be introduced in the linking process due to missed-matches or false-matches. Missed-matches are records which belong to the same individual or entity but fail to be matched. False - matches are records that are erroneously matched but belong to two different individuals or entities. These two possible linkage errors can produce biased estimate. Gomatam et al. (2002), Nitsch et al. (2006) and Fair et al. (2000) show that bias increases as linkage error increases. It is difficult to measure the extent of this bias with the formal measures of linkage errors such as sensitivity, specificity or match rate (Bohensky et al., 2010; Christen & Goiser, 2005; Leiss, 2007; Neter et al., 1965; Campbell, 2009). Belin and Rubin (1995) suggest 'false match rate', which is estimated by, "$1 - $ (number of true links/total links)". They consider the distribution of observed weights as a mixture of two distributions. One distribution is for matched pairs and the other for non-matched pairs and these distributions are estimated by fitting transformed normal curves to the record pair weights. Winkler (2007) noted that their method performs well when the curves are well separated for matches and non-matches. The input parameters estimates for the mixture model depends on the quality of training data and without a good training data these estimates will affect the estimated error rates. Winkler (2007) provided an unsupervised learning method without training data for automatically estimating false match rates. Winglee, Valliant and Scheuren (2005) designed a simulation approach, *Simrate* to estimate error rates. Their method uses the observed distribution of data in matched and non-matched pairs to generate a large simulated set of record pairs. They assign a match weight to each record pair following specified match rules and use the weight distribution for error estimation. Larsen & Rubin (2001) estimate true match status by using the posterior probability of a match and improve the classification of matches and non-matches through clerical review. However, clerical review can be expensive and time consuming. Moreover, it is not possible to identify a correct link even after the clerical review. Winkler and Yancey (2006) demonstrated two subsets of pairs, one is 'pseudo-true' matches and another 'pseudo-true' non-matches. Record pairs are considered as 'pseudo-true' matches above a certain score and 'pseudo-true' non-matches below another lower score. Using this artificial 'pseudo-truth' set under various modelling assumptions accurate estimates of false match and non-match rates can be obtained using non 1-1 matching in limited situations. In general, however their method does not work because it is not always possible to separate set of matches from non-matches.

Fortini and others (2001) proposed a Bayesian approach for record linkage by using standard Metropolis-Hastings and Simulated Annealing algorithms to derive the marginal distribution. Missing links or false negatives can make datasets unrepresentative of the total population of true links. To address this problem, McGlincy (2004) developed a full Bayesian model for the posterior probability that a record pair is a true match given observed agreements and disagreements of comparison fields. This method gives representative sets of imputed linked record pairs. Liseo et al. (2011) outlined the record linkage issue into a formal inferential problem and improved standard model selection techniques by reviewing recent advances on Bayesian methodology. Goldstein et al. (2012) argued that existing methods that select record that have the maximum weight larger than an assigned threshold are inefficient and may also lead to biases in subsequent analysis of the linked data as it ignores all information from matches with lower weights and for some individuals assigns no match. They proposed a multiple imputation framework to obtain information from all potential matches at the analysis stage. Their simulation results suggest multiple imputation to obtain unbiased and efficient parameter estimates rather than go to a full probabilistic linkage. Sayers et al. (2016) illustrated the critical steps involving probabilistic record linkage using a simple exemplar. They described the process of calculating matched weights and converting matched weights into posterior probabilities of a match using Bayes theorem. While emphasizing the benefit of using probabilistic record linkage to improve the linkage, they also noticed the complex issues of current research in record linkage field including privacy (Smith et al. 2014), analysis of linked data sets (Goldstein et al. 2012) and automated selection of matched and non-matched records using EM algorithm (Grannis et al. 2003).

One way of measuring linkage error is by the proportion of links that are correct matches. Incorrect links create measurement error and bias the analysis (Harron et al. 2014; Chipperfield et al. 2011; Chipperfield and Chambers 2015; Chambers et al. 2009; Lahiri and Larsen 2005). Harron et al. (2014) assessed the impact of linkage error on estimated infection rates in paediatric intensive care based on linking a national audit dataset (PICA - Net, the Paediatric Intensive Care Audit Network) and infection surveillance data (Paediatric Intensive Care Audit Network National Report, 2009 – 2011). Their study found that the bias was greater when the match rate was low or the error rate in variable values was high. In their analysis, they assumed that both the match weight and match probabilities were calculated accurately as they are based on the true match status of record pairs. However, this would not be the case in real linkage situation. Resnick and Asher (2019) proposed a new Type I and Type II error measurement technique where they developed synthetic data with known match patterns, apply probabilistic matching process on this data. The results of the probabilistic match process are then compared to the known match pattern to estimate error. The error measurement technique needs further exploration to determine which record linkage conditions are required for its use.

In regression analysis, the possible linkage error affects the estimation of the relationships between variables of the two files. Neter et al. (1965) first noted that the presence of 'false matches' reduces the observed level of association between variables. Later, Scheuren and Winkler (1993, 1997) and Lahiri and Larsen (2005) discussed this problem in detail. They introduce bias when estimating the slope of the regression line. Lahiri and Larsen (2005) and Scheuren and Winkler (1993) proposed methods to calculate unbiased estimates of coefficients for a linear regression model given data from a probabilistically linked file. Later, Chambers, Chipperfield, Davis and Kovacevic (2009) and Chambers (2008) proposed models using generalised estimating equations when linking two files where one file is a subset of the other file. Chipperfield et. al (2011) analyses binary variables. More recently, Kim and Chambers (2012, 2013) extended this work to a wide set of models using estimating equations. Di Consiglio et al. (2018) provides a sensitivity analysis of the effect of linkage errors both on bias and variability of regression estimates. In their analysis linkage errors are assumed to be known. They showed that the correction for the bias is more effective in the linear than in the logistic model and missing matches should be considered to completely remove the bias. They suggest an assessment of the trade-off between the adjustment of the bias and the expected increase in variance while estimating linkage errors.

Chipperfield and Chambers (2015) developed a parametric bootstrap method of making inferences for binary variables using probabilistic method to link file under the 1-1 constraint. They showed that the analytic estimates of precision in Lahiri and Larsen (2005) are poor for 1-1 probabilistic linkage. Lahiri & Larsen (2005) do not consider 1-1 linkage where every record from one file is linked to a distinct record in another file. Chipperfield and Chambers's (2015) method is valid as long as the linkage process can be replicated.

So far, we have discussed some of the contributions in the field focusing on the quality of the linkage. We continue our discussion particularly on different comparators used for comparison of records. Record linkage involves comparison of records, looking for agreement and disagreement of values between files. There are different methods in which this comparison is undertaken. Since this paper is focused on similarity examination between records, this motivates the following discussion about various comparators of values including strings.

To improve matching efficacy, Newcombe et al. (1959, 1962) introduced the idea of relative frequency of strings. However, Winkler (1989c) and Yancey (2000) noted that in a number of situations the use of relative frequency will not improve matching. The problem of using the relative frequency of strings rigorously is that in some cases the two files may not overlap, and then agreement on a rare value for a field has more discriminatory power in matching than agreement on a frequently occurring value. Also, higher typographical error rates for rare names can result in misleading relative frequency tables.

The computer science literature proposes some advanced methods for record linkage based on textual information, such as, name and address. Some methods focus on parsing and standardising names and addresses into components so that they can be easily compared (e.g., Winkler 1995, Borkar et al. 2001, Christen et al. 2002, Churches et al. 2002). Parsing is the process of separating a sentence into grammatical parts and standardization involves replacing various spelling of words with a single spelling. Name standardization means identifying components such as first names, titles, surnames or last names etc. and address standardization determines components, for example, house numbers, street numbers, post office boxes etc. Linking records for the variants of names and addresses is not easy, especially for business lists (Winkler 1995).

Winkler (1990a) and Cohen et al. (2003a,b) proposed methods for string comparison based on typographical error. For many typographical errors, it is not always possible to compare two string fields character-by-character. String comparison aims to compare pairs of strings, for example, 'Smith' and 'Snith'. Approximate string comparison has been an important research area for many authors in the computer science literature (e.g., Hall and Dowling 1980, Navarro 2001). The function for the approximate agreement takes values between 0 and 1 by allowing degrees of partial transpositions. The complete agreement of values is represented by 1 and the total disagreement by 0.

Jaro (1976, see also 1989) proposed a string comparator that takes a value of 1 for exact agreement and takes values less than 1 for partial agreement. The algorithm uses the string lengths, the number of common characters in the two strings and the number of transpositions. The transpositions represent the number of characters in one string that are not in the correct sequence of the corresponding common character from the other string. The distance of the common characters (agreeing character) between strings is half the length of the shorter string.

Winkler (1990a) used truth data sets (when data in the form of known true matches and non-matches are available) to model the effect of different values of string comparators on the likelihood in the Fellegi-Sunter decision rule. Winkler showed that the alternative to the Jaro string comparator improves matching compared to a no string comparator. Winkler's method uses some ideas of Pollock and Zamora (1984) where empirical evidence of the increasing probability of keypunch errors is provided. Keypunch errors occur when the character changes position, e.g., moves to the right in a string. Merging administrative files from a variety of sources is necessary to facilitate matching. Agencies such as Eurostat connect data from different sources and countries for merging data files. For merging purposes, agencies use software for standardizing and cleaning lists. For comparison of records, standardization software breaks name and addresses into components. The matching software acts on typographical error and automatically estimating matching parameters. Winkler (2001) describes methods and software that deals with cleaning and standardizing lists. He also pointed on the substantial amounts of skill and time required for data processing to improve matching.

The methods mentioned above may not perform well in case of large numbers of typographical errors or non-standardization of records. The inconsistencies of name and address information, non-homogeneous identifying characteristics of record pairs, lack of easy comparable variables for matching, missing matching variables and so on may make automatic matching infeasible or impossible.

As massive amounts of data are now being collected by organizations in the private and public sectors, the requirement of linked data from different sources is also increasing. Linking and analysing relevant entities from various sources provide benefits to businesses and government organizations. However, linking variables may not identify a person. Because real world data sometimes contain errors, they may change over time or missing. Data entry tends to contain errors, especially in the case of names and addresses when they are received over the phone, hand-written or scanned. Name and address may also change over time. Typographical errors or spelling errors and different formats of data also sometimes make it hard to identify a correct link. Although, a lot of work has been done which measures the quality of the linked file, less attention has been paid to methods for assessing the linking method.

In our previous work, we proposed a Markov Chain based Monte Carlo simulation approach (*MaCSim*) for assessing the linking method. *MaCSim* utilizes two linked files that have been previously linked on similar types of data to create an agreement matrix and obtain observed links using a defined linking method. The agreement matrix is simulated using a simulation algorithm to obtain simulated links using the same linking method. The approach calculates the accuracy of the linking method based on correct relink proportions by comparing the observed and simulated links.

In this paper, we add an extra feature to our original *MaCSim* method. The aim is to improve the assessment of accuracy of linkages. We calculate a similarity weight for every linking variable value, which allows partial agreement of the linking variable values for record pairs. The extended *MaCSim* uses similarity weights to create the agreement matrix. The agreement matrix is simulated using the adjusted *MaCSim* algorithm to generate samples. Like the original *MaCSim*, observed links are obtained using a defined linking method and are compared with simulated links to assess the linking method based on correct relink proportions.

The paper is organised as follows. Section 2 provides a description of the original *MaCSim*. The topic moves to the extended *MaCSim* algorithm in Section 3. In Section 4, we present results of application of the new algorithm to a synthetic dataset provided by the Australian Bureau of Statistics (ABS) based on realistic data settings. The paper concludes with a summary in Section 5.

## 2. Original *MaCSim* approach

Before moving to the discussion of the extension of *MaCSim*, this Section gives an overview of the original *MaCSim* approach (Haque et al. 2020, submitted).

### 2.1 The purpose of developing *MaCSim*

When there is a task to link two files, it is hard to decide which method to use for linking. Since these are new files, there is no way to measure the accuracy after linking without further review. *MaCSim* can assist in the evaluation of which method will give higher accuracy to link these files. The approach can be used as a tool to assess a linking method, or to evaluate or compare other linking methods. Based on the obtained accuracy results, the user can decide on a preferred method or evaluate whether it is worth linking the two files at all.

## 2.2 *MaCSim*

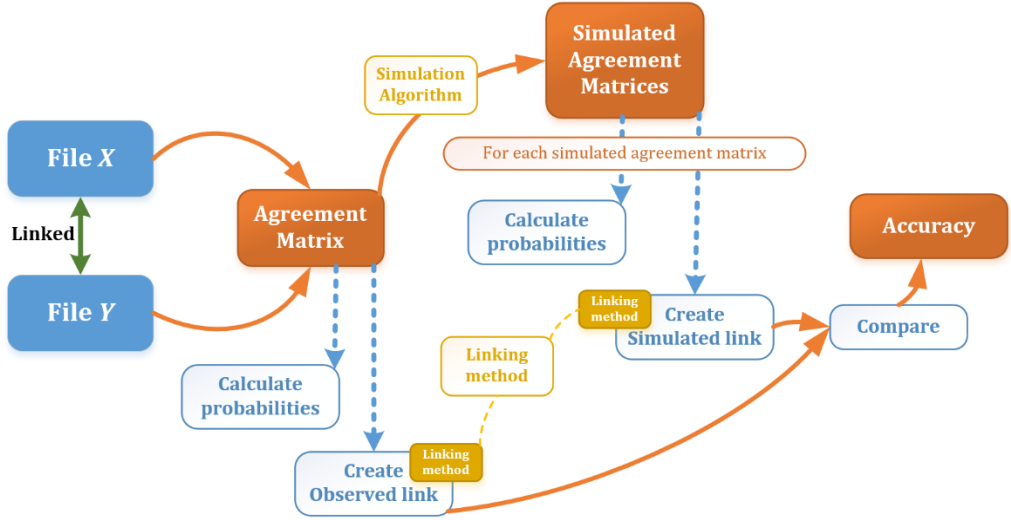

**Figure 1: *MaCSim***

In our previous work, we developed a Markov Chain based Monte Carlo simulation method (*MaCSim*) for assessing a linking method. As we see in the Figure 1, *MaCSim* utilizes two linked files to create an agreement matrix. From this agreement matrix the necessary parameter values are calculated and links are constructed using a defined linking method. The agreement matrix is then simulated using a defined algorithm developed for generating re-sampled versions of the agreement matrix. In each simulation with the simulated data, records are re-linked using the same linking method. The simulated link is then compared with the observed link and the accuracy of the individual link is calculated. This ultimately provides an evaluation of the accuracy of the linking method that has been followed to link the records.

### 2.3 Creating agreement matrix *A*

An agreement matrix, ***A***, is created from the two files to be linked, *X* and *Y*, where

$$\boldsymbol{A} = (A_{ijl}); \quad i = 1, \dots R_X, \quad j = 1, \dots R_Y, \quad l = 1, \dots, L,$$

is a three-dimensional array denoting the agreement pattern of all linking variables across all records in the two files. Here, $A_{ijl} = 1$ if the *lth* linking field value for record *i* of file *X* and record *j* of file *Y*, are the same; $A_{ijl} = -1$ if these values are not the same and $A_{ijl} = 0$ if either or both the values are missing. Therefore, an agreement matrix contains agreement values 1, -1, and 0, which are the comparison outcome between record pairs of the two files to be linked.

For simplicity of notation we assume that $A_{iil}$ represents the agreement value of the $lth$ linking variable for the true matched record pair in both files.

## 2.4 Simulating agreement matrix *A*

To assess standard errors for estimates deriving from analysis of the linked data, our interest is to generate re-sampled versions of the agreement matrix **A** in a way so that it preserves the underlying probabilistic linking structure. For this purpose, the *MaCSim* algorithm develops a Markov Chain $\{A^{(n)}\}_{n=0,1,2,...}$ on *A*={set of possible agreement pattern arrays}, with $A^{(0)} = A$, the observed agreement pattern array for the files *X* and *Y*. The key step is to simulate the observed agreement matrix **A** to create $A^*$ which includes all the simulated agreement matrices.

## 2.5 Original *MaCSim* simulation algorithm

The structure of the transition probabilities for the MCMC algorithm employed by *MaCSim* is outlined. Given the current state of the chain, $A^{(n)}$, the next state, $A^{(n+1)}$, is constructed as follows:

*Step 1*: Initially, set $A_{ijl}^{(n+1)} = A_{ijl}^{(n)}$ for all $i, j,$ and $l$.

*Step 2*: Randomly select values of $i \in \{1, ..., R_X\}$ and $l \in \{1, ..., L\}$.

*Step 3*: If

    a) $A_{iil}^{(n)} = 1$, change $A_{iil}^{(n+1)}$ to –1 with probability $p_1$.

    b) $A_{iil}^{(n)} = -1$, change $A_{iil}^{(n+1)}$ to 1 with probability $p_2$.

*Step 4*: For each $j \neq i$, if

    a) $A_{iil}^{(n)} = 1$ & $A_{iil}^{(n+1)} = -1$, then

        i) If $A_{ijl}^{(n)} = 1$, change $A_{ijl}^{(n+1)}$ to –1.

        ii) If $A_{ijl}^{(n)} = -1$, change $A_{ijl}^{(n+1)}$ to 1 with probability $q_1$.

    b) $A_{iil}^{(n)} = -1$ & $A_{iil}^{(n+1)} = 1$ then

        i) If $A_{ijl}^{(n)} = 1$, change $A_{ijl}^{(n+1)}$ to –1.

        ii) If $A_{ijl}^{(n)} = -1$, change $A_{ijl}^{(n+1)}$ to 1 with probability $q_2$.

c) $A_{iil}^{(n)} = -1$ & $A_{iil}^{(n+1)} = -1$ then

If $A_{ijl}^{(n)} = -1$, change $A_{ijl}^{(n+1)}$ to 1 with probability $q_3$.

The values of $p = (p_1, p_2)$ and $q = (q_1, q_2, q_3)$ are determined to ensure the stationary distribution of the chain has the desired structure. Thus, this Markov chain can be used to generate an appropriate set of re-sampled **A** values.

## 2.6 Maintaining internal agreement consistency

The transition structure is designed to replicate the circumstances whereby a random element of file $X$ is selected and then a change in its value for the $l$th linking variable is made with probability based on its current agreement status with its corresponding partner in the opposite file. If a change does occur, this will have a consequent effect on the agreement patterns in the associated non-matching record pairs. For the matched record pair, if the agreement value of the selected linking variable is changed, that is, from 1 to -1 or -1 to 1 with probability $p_1$ and $p_2$ as in steps $3(a)$ and $3(b)$, then this will have a consequent effect on the agreement patterns in the associated non-matching record pairs. Therefore, if the agreement value of the associated non-matching record pair is 1 then we must change it to -1 as they can no longer agree ($4ai$ & $4bi$). However, if the agreement value of the associated non-matching record pair is -1, then we change it to 1 with probability $q_1$, $q_2$ and $q_3$ because now the value may or may not agree ($4aii, 4bii$ & $4c$). With this underpinning, the internal consistency patterns of agreement will be maintained.

## 2.7 Maintaining marginal distributions

In addition to internal agreement consistency, the chain maintains the required probabilities of agreement for both matched and non-matched records across the two files. This requires appropriate selection of the transition probability parameters $p = (p_1, p_2)$ and $q = (q_1, q_2, q_3)$. In particular, we require the probability that linking field values for matched record pairs agree remains $m_l$. That is, $Pr\{A_{iil}^{(n+1)} = 1\} = m_l$. Choosing appropriate values for the $q$ parameters arises from the requirement to maintain the probability of agreement between values of the linking variable among non-matched records. In other words, we must ensure that $Pr\{A_{ijl}^{(n+1)} = 1\} = u_l$.

To maintain the marginal probabilities of matching, we choose the transition probability parameters as follows:

$$p_1 = \begin{cases} (1 - m_l - g_l)/m_l & \text{if } u_l \leq 0.5(1 - g_l) \\ (1 - m_l - g_l)(1 - u_l - g_l)/\{m_l(3u_l + g_l - 1)\} & \text{otherwise} \end{cases}$$

$$p_2 = p_1 \, m_l/(1 - m_l - g_l)$$

$$q_1 = q_2 = \begin{cases} u_l/(1 - u_l - g_l) & \text{if } u_l \leq 0.5(1 - g_l) \\ 1 & \text{otherwise} \end{cases}$$

$$q_3 = 1.$$

The detailed derivation of these values is provided in our previous paper (Haque et al. 2020, submitted).

## 2.8 Calculating $m, u$ and $g$ probabilities

To recap, $m$ is the probability that the variable values agree when the record pair represents the same entity; $u$ is the probability that the variable values agree when the record pair represents two different entities, and $g$ is the probability when the variable values are missing from either or both records in the pair.

For each linking variable $l$, $m_l$, $u_l$ and $g_l$ are calculated in the following way:

$\widehat{m_l}$ = number of values that agree for matched record pairs/total number of matched record pairs.

$\widehat{u_l}$ = number of values that agree for non-matched record pairs/total number of non-matched record pairs.

$\widehat{g_l}$ = total number of record pairs of which one or both values are missing/total number of possible record pairs.

## 2.9 Creating observed link

After comparing each linking variable value for a record pair from the two files, the result is a ternary code, 1 (when values agree), -1 (when values disagree) and 0 (when either or both values are missing) in the agreement matrix $A$. According to these codes, each linking variable is given a weight using the probabilities $m_l, u_l,$ and $g_l$. For any $(i, j)$-th record pair and any linking variable $l$, if the agreement value is 1 (i.e. $A_{ijl}=1$) then the weight is calculated using $w_{ijl} = log\left(\frac{m_l}{u_l}\right)$; if the value is -1 (i.e. $A_{ijl}=-1$), the weight is calculated using $w_{ijl} = log(1 - m_l - g_l)/(1 - u_l - g_l)$ and for a missing value (i.e. $A_{ijl}=0$), the weight formula is $w_{ijl} = log\ (g_l/g_l) = log\ (1)$.

Since we assume that missingness occurs at random, and therefore has the same chance of occurring in a true matched pair as in a non-match, missing values will not contribute to the weight.

After calculating the weight for each record pair that agree or disagree on a linking variable value explained above, a composite or overall weight, $W_{ij}$ is calculated for each record pair $(i, j)$ by summing individual weights, $w_{ijl}$ over all linking variables $l$ for that pair using the following formula:

$$W_{ij} = \sum_l w_{ijl}$$

Once weights of all record pairs, $W_{ij}$ are calculated, the observed links are created following the steps of defined linking method below:
a. First, all record pairs are sorted by their weight, from largest to smallest.
b. The first record pair in the ordered list is linked if it has a weight greater than the chosen cut-off value.
c. In all the other record pairs that contain either of the records from the associated record pair that have been linked in step b, are removed from the list. Thus, possible duplicate links are discarded.
d. Go to step b for the second record and so on until no more records can be linked.

To calculate the accuracy of the linking method above, these observed links are compared with simulated links obtained by using the same linking method. It is worth mentioning that *MaCSim* does not consider consistency of accuracy through simulations to assess a linking method. It measures the average accuracy of each record in all simulations and average accuracy for all records in every simulation. Based on the obtained accuracy results, the user can choose among methods for the linking task or evaluate whether it is worth linking.

The aim of *MaCSim* is to assess a linking method. It needs two linked files that have been previously linked on similar types of data to ultimately assess or help decide which linking method to use for linking new files or even whether it is worth linking the files. Match and non- match probabilities can be estimated using a linked file or they may be known from previous linkages of similar types of data. The main purpose of *MaCSim* is not to link new files. However, the linking method used in *MaCSim* could be a better choice to link files; indeed, it has already been tested on a dataset and gave high accuracy.

## 3. *MaCSim* with similarity weight

In this paper, we extend the original *MaCSim* to improve the assessment of the accuracy of record linkage. The extended *MaCSim* uses partial agreement of the linking variable values in the form of a similarity weight. A similarity weight is calculated for every linking variable value for each record pair. In the linking process, a record pair will be considered to agree on a linking variable if its similarity weight exceeds the corresponding threshold. A threshold is calculated for every linking variable to decide on an acceptable difference in values, based on a pre-defined tolerance for that variable. Since it is not practical to set the same tolerance limit for agreement or disagreement of values for different linking variables, the user can choose a different "tolerance" for each linking variable. "Tolerance" is a parameter that user can specify based on the variable of interest and thus the user can choose a different agreement threshold for different values of tolerance for each linking variable. For example, the user can set a tolerance value of 5 for age difference and 0 for eye color. In our case study, variable values are coded by numbers. For example, five categories of eye color are numbered from 1 to 5.

To assess the linking method, the extended *MaCSim* employs the same steps as the original *MaCSim* (Section 2): create an agreement matrix; generate re-sampled versions of the agreement matrix; employ a defined linking method to link records using the simulated agreement matrices; and calculate the linkage accuracy for every record in each simulation. These steps of the extended *MaCSim* with additional feature are described in the following sections.

### 3.1 Create agreement matrix *A*

An agreement matrix, *A* is created with similarity weight for each record pair for every linking variable from the two linked files, $X$ and $Y$, where

$$A = A_{ijl}; \quad i = 1, \dots R_X, \quad j = 1, \dots R_Y, \quad l = 1, \dots, L,$$

is a three-dimensional array consisting of similarity weights of all linking variables across all records in the two files. Here, $A_{ijl}$ takes real values in the range 0 to 1 according to the partial agreement of linking variable $l$ for record pair $(i,j)$. The partial agreement values are given using a defined similarity weight formula (Section 3.2). Values of 0 and 1 indicate complete disagreement and agreement, respectively. When either or both of the values in the $l$th linking variable of file $X$ and file $Y$ are missing, a value of -1 is given.

### 3.2 Similarity weight calculation

The similarity weight of any record pair $(i, j)$ for every linking variable $l$ is calculated by the formula,

$$V_{ijl} = 1 - {d_{ijl}}/{T_l}$$

where, for any record pair $(i, j)$, $d_{ijl}$ is calculated by the difference between the values for the variable $l$. $T_l$ is the difference between the maximum and minimum values of variable $l$. We set a tolerance for each variable $l$, i.e. $\delta_l$ which is the maximum difference of values of the linking variable $l$ that can be accepted as an agreement. We calculate an individual agreement threshold for each linking variable $l$ by

$$\theta_l = 1 - {\delta_l}/{T_l}$$

For any record pair $(i, j)$ and linking variable $l$, if the similarity weight $V_{ijl}$ is greater than or equal to the value of agreement threshold $\theta_l$, then the variable value is accepted as an agreement.

The user can choose a different agreement threshold for different values of tolerance for each linking variable.

## 3.3 Simulating the agreement matrix *A*

Like the original *MaCSim*, the observed agreement matrix **A** is simulated using a defined algorithm, described below to create **A***, where **A*** contains all the simulated agreement matrices. The original *MaCSim* algorithm described in Section 2.5 is modified here to reflect the changes of agreement matrix with partial agreement values.

The structure of the transition probabilities for the MCMC algorithm employed by the extended *MaCSim* is outlined. Given the current state of the chain, **A**$^{(n)}$, the next state, **A**$^{(n+1)}$, will be constructed as follows:

*Step 1*: Initially, set $A_{ijl}^{(n+1)} = A_{ijl}^{(n)}$ for all $i, j,$ and $l$.

*Step 2*: Randomly select values of $i \in \{1, ..., R_X\}$ and $l \in \{1, ..., L\}$.

*Step 3*: If

a) $A_{iil}^{(n)} \geq \theta_l$, change $A_{iil}^{(n+1)}$ to $(1-A_{iil}^{(n+1)})$ with probability $p_1$.

b) $A_{iil}^{(n)} < \theta_l$, change $A_{iil}^{(n+1)}$ to $(1-A_{iil}^{(n+1)})$ with probability $p_2$.

*Step 4*: For each $j \neq i$, if

a) $A_{iil}^{(n)} \geq \theta_l$ & $A_{iil}^{(n+1)} < \theta_l$, then

i) If $A_{ijl}^{(n)} \geq \theta_l$, change $A_{ijl}^{(n+1)}$ to $(1-A_{ijl}^{(n+1)})$.

ii) If $A_{ijl}^{(n)} < \theta_l$, change $A_{ijl}^{(n+1)}$ to $(1-A_{ijl}^{(n+1)})$ with probability $q_1$.

b) $A_{iil}^{(n)} < \theta_l$ & $A_{iil}^{(n+1)} \geq \theta_l$ then

i) If $A_{ijl}^{(n)} \geq \theta_l$, change $A_{ijl}^{(n+1)}$ to $(1-A_{ijl}^{(n+1)})$.

ii) If $A_{ijl}^{(n)} < \theta_l$, change $A_{ijl}^{(n+1)}$ to $(1-A_{ijl}^{(n+1)})$ with probability $q_2$.

c) $A_{iil}^{(n)} < \theta_l$ & $A_{iil}^{(n+1)} < \theta_l$ then

If $A_{ijl}^{(n)} < \theta_l$, change $A_{ijl}^{(n+1)}$ to $(1-A_{ijl}^{(n+1)})$ with probability $q_3$.

The extended *MaCSim* considers the initial agreement matrix as the current state, and follows the algorithm to determine the next state that is the next simulated agreement matrix. In the next iteration, the new simulated agreement matrix becomes current state, and in the same way following the algorithm it determines the next state that is another instance of agreement matrix and so on.

Similar to our original *MaCSim* algorithm (Section 2.5), the structure of the transition probabilities for the proposed Markov chain is outlined to replicate circumstances by which a random element of file $X$ is selected and then its value for the $lth$ linking variable is changed with probability based on its current agreement status with its corresponding partner in the opposite file. On this basis, the internal consistency patterns of agreement are maintained. If a change has been made, this has the consequent effect of changing the agreement patterns in the associated non-matching record pairs. The algorithm describes all possible cases for which a change in one variable value of match and non-match records has an impact of other record pairs and changes values accordingly. In the algorithm, the agreement and disagreement values for matched and non-matched records (whether a record pair agrees or disagrees on a variable) are compared to individual threshold of the variable. Note that, in the original *MaCSim* algorithm, it was compared to 1 (for agree) or -1 (for disagree). Based on this comparison, an agree or disagree decision is reached. The current state of the algorithm determines whether the agreement or disagreement value will be changed with probability in the next state. Previously, the changes are made by '1 to -1', and '-1 to 1'. In the extended *MaCSim*, if we decided to change the value in the next state, we change it to (1-value in current state). Therefore, the extended *MaCSim* algorithm also maintains internal agreement consistency and the values of the transition probability parameters, $p = (p_1, p_2)$ and $q = (q_1, q_2, q_3)$, are selected to ensure that the stationary distribution of the chain maintains the required probabilities of agreement for both matched and non-matched records across the two files (Section 2.7). Once values for $p$ and $q$ are determined to ensure that the stationary distribution of the chain has the desired structure, this Markov chain is used to generate an appropriate set of re-sampled agreement matrices.

### 3.4 Calculate $m, u$ and $g$ probabilities

For each linking variable, $m_l$, $u_l$ and $g_l$ are calculated in the following way:

$\widehat{m_l}$ = number of values that agree for matched record pairs/total number of matched record pairs.

$\hat{u}_l$ = number of values that agree for non-matched record pairs/total number of non-matched record pairs.

$\hat{g}_l$ = total number of record pairs of which one or both values are missing/total number of possible record pairs.

### 3.5 Create observed link

To create the observed links, the weight of each record pairs is calculated from the agreement matrix $A$ using the probabilities $m_l$, $u_l$ and $g_l$. For any record pair $(i, j)$ and any linking variable $l$, the weight is calculated by

$$w_{ijl} = log\left(\frac{m_l}{u_l}\right)$$

when the values agree (i.e. $A_{ijl} \geq \theta_l$); if the values disagree (i.e. $A_{ijl} < \theta_l$), the weight is calculated using

$$w_{ijl} = log(1 - m_l - g_l)/(1 - u_l - g_l)$$

and for a missing value (i.e. $A_{ijl}$=-1), the weight formula is

$$w_{ijl} = log\ (g_l/g_l) = log\ (1).$$

A composite or overall weight, $W_{ij}$ is calculated for each record pair $(i, j)$ by summing individual weights, $w_{ijl}$ over all linking variables $l$ for that pair using the following formula:

$$W_{ij} = \sum_l w_{ijl}$$

After all the weights $W_{ij}$ of all record pairs are calculated, the observed links are created following the steps below:

I. First, all record pairs are sorted by their weight, from largest to smallest.
II. The first record pair in the ordered list is linked if it has a weight greater than the chosen cut-off value.
III. All the other record pairs containing either of the records that have been linked in step b are removed from the list. Thus, possible duplicate links are discarded.
IV. Go to step II for the second record and so on until no more records can be linked.

## 4. Analysis

### 4.1 Analysis setup

For the analysis, a synthetic dataset is used which consists of information regarding hypothetical individuals. The dataset has been generated by the Australian Bureau of Statistics (ABS). A large file $Y$ is created with 400,000 records. Then a random subsample of 50,000 records is taken from file $Y$ to form file $X$. Every record has eight data fields or linking variables. These variables are: RECID (Record Identifier), SA1 (Statistical Area 1), MB (Meshblock), BDAY (Birth day), BYEAR (Birth year), SEX (Male/Female), EYE (Eye colour) and COB (Country of birth). The value of each linking field (or linking variable) is generated independently except the value of COB. For COB, 300,000 records are coded by '1101' indicating 'Born in Australia' and the remaining 100,000 records are randomly assigned one of about 300 country codes according to the corresponding proportion of people in the 2006 Australian Census.

As part of the data preparation, some values in file $X$ are changed by replacing them either with a randomly chosen value from the records in file $Y$ or setting the value to 'missing' to simulate errors in linking variables. For this modification, individual records are selected independently. The value of SA1 variable is changed to an adjacent SA1 for 500 (1%) records, and the first five digits of the corresponding Meshblock code are altered appropriately. The MB is changed to another MB within the same SA1 region for 1,500 (3%) records. BDAY is changed to 'missing' for 4,000 (8%) records. For 500 records (1%), the day and month corresponding to the numeric code are altered. In the BYEAR field, 50 records are replaced with 'BYEAR–2', 50 with 'BYEAR+2'. 1200 records are reset to 'BYEAR–1' and 1200 to 'BYEAR+1'. For the SEX variable, the value of 50 records (0.1%) is reversed. The value of EYE variable is set to 'missing' for 5,000 records (10%). For another 5,000 records (10%), a valid alternative is chosen as a replacement value. The COB variable is set to 'missing' for 750 records (approximately 2%) of the records coded to "1101". COB is also set to 'missing' for 250 records (approximately 2%) with another country code. For 125 of these cases, records are replaced with 'Australia' and for the remaining 125 cases, records in COB are recoded to another country within the same broad geographical region (e.g. with the same two-digit SACC code) (Peter Rossiter, 2014). In both files, the RECID (Record Identifier) stays the same. This makes it easy to identify true matches and non-matches in the linking process. RECID is not used for linking or blocking.

**4.2 Analyses results**

Two different blocking variables, namely SA1 (Statistical Area 1), and combined variable SA1 & SEX, are used for the analyses. Blocking takes into account those record pairs in a block to compare that have the same value for a blocking variable, thus reducing the number of comparisons of record pairs.

The initial agreement matrix $\boldsymbol{A}$ is simulated to obtain $S = 1{,}000$ replicates $\boldsymbol{A}^*$ of $\boldsymbol{A}$. In particular, we can set $\boldsymbol{A}^{*(s)} = \boldsymbol{A}^{(sd)}$, for $s = 1, \ldots, S$ and some constant $d$. By the nature of the MCMC process, the elements of the sample can be highly correlated. The "thinning" parameter allows us to specify whether and by how much the MCMC chains should be subsampled in order to reduce this correlation. In our case, the value of the thinning parameter $d$ is set to 1,000 which results in keeping every 1000th value and discarding all interim values. Therefore, 1,000,000 MCMC simulations are run and $s$ samples $\boldsymbol{A}^{(s)}$, where, $s = 1, \ldots, 1000$, are retained. Thus, in $\boldsymbol{A}^*$, we have 1000 instances of the agreement matrix $\boldsymbol{A}$.

**4.2.1 Distances**

The distances between the entries of each $\boldsymbol{A}^*$ ($\boldsymbol{A}^{*(2)}$, $\boldsymbol{A}^{*(3)}$, ...., $\boldsymbol{A}^{*(S)}$) and the initial agreement matrix $\boldsymbol{A}^{*(1)}$ are calculated. In every simulation, the distance is calculated by the total number of agreement values that are changed from the initial values divided by the total number of agreement values. The distance plot in Figure 2 shows the proportion of agreement values that are changing in each simulation.

Figure 2 shows the distances in 1000 simulations using two blocking variables: (i) blocking variable SA1, (ii) combined variable SA1 & SEX.

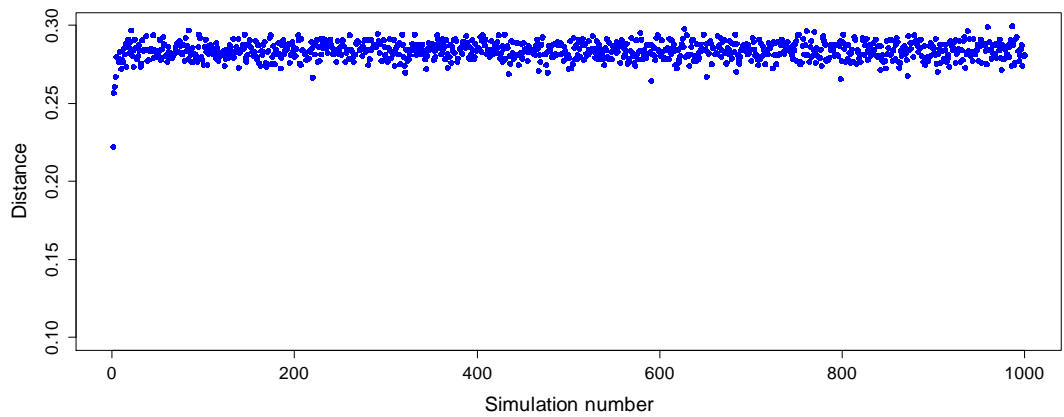

**(i) Distance with blocking variable SA1**

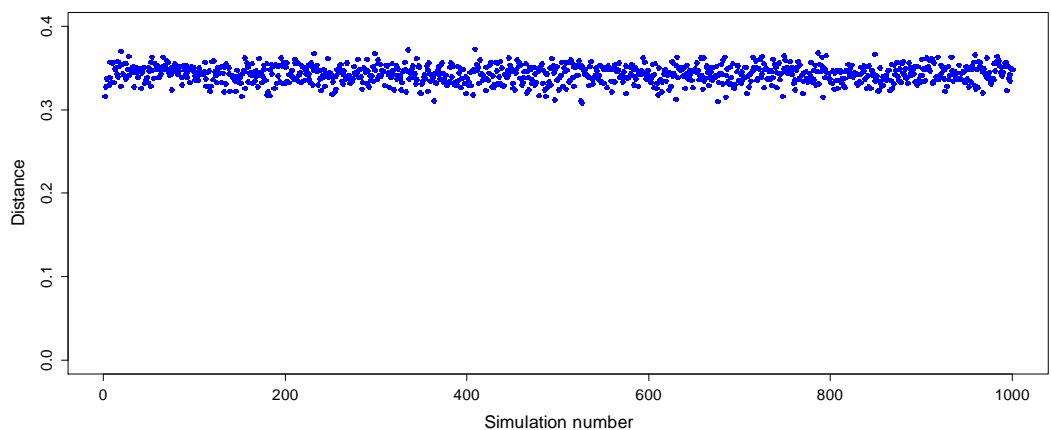

**(ii) Distance with blocking variable SA1 & SEX**

**Figure 2: Distance of $A^*$ entries from the initial agreement matrix**

From the distance plot on blocking variable 'SA1' in Figure 2 (i), approximately 28% of the values in the elements of $A^*$ are changed. The chain converged after 10 iterations. The chain stays stable in 1000 simulations. For the combined blocking variable SA1 & SEX, the plot (Figure 2 (ii)) shows that the chain converges after 5 iterations to around 0.35. Note that convergence occurs at two different points for these two blocking variables, SA1 and SA1 & SEX. For both variables, the proportions of agree, disagree, and missing values are different and in each simulation the chances of agreement and disagreement in the next state depend on the respective values in the current state. In MCMC sampling, once the chain has converged, its elements can be seen as a sample from the target distribution. The distance plots (Figure 2) for both single and combined variables show convergence of the chain.

### 4.2.2 Correct re-link proportion per X record

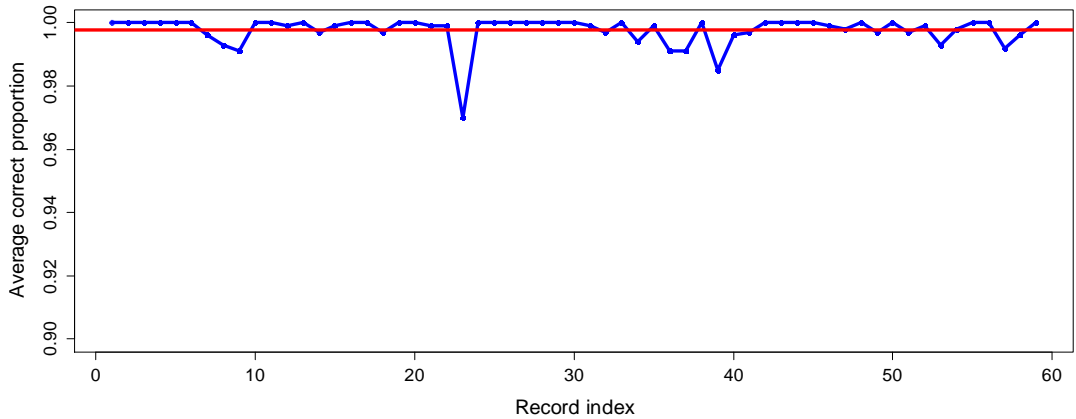

**(i) Correct re-link proportion with blocking variable SA1**

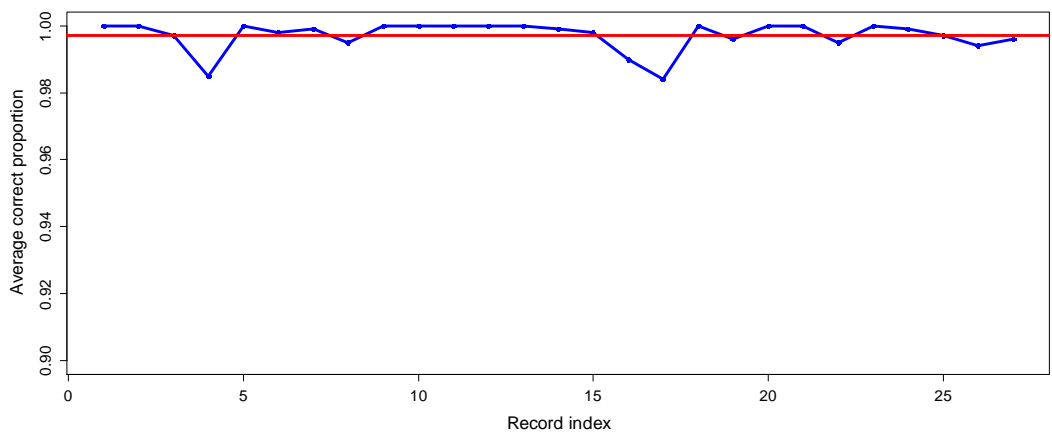

**(ii) Correct re-link proportion with blocking variable SA1 & SEX**

**Figure 3: Correct re-link proportion of each X record**

In every simulation, the records are linked using the simulated agreement matrices in $A^*$, following the defined linking method in Section 3.5. We observe the total number of times each record is re-linked to the record to which it was originally linked.

When blocking the data with SA1, the first block contains 59 records in File $X$. Figure 3 (i) shows the correct re-link proportion of each X record in 1000 simulations. The plot shows that the proportion of correct re-links for each of the 59 records lies between 97% and 100% and the average accuracy is 99.75%, indicated with the red line. The error for each individual record is low with a maximum error of 3% for record number 23.

With the combined variable SA1 & SEX (Figure 3 (ii)), there are 26 records in the first block in file *X*. The correct re-link proportion of each X record exceeds 98% in the 1000 simulations. In the plot, the red line shows that the average accuracy is 99.8%. The maximum error is only 1.8%, for record number 17.

### 4.2.3 Correct re-link proportion per simulation

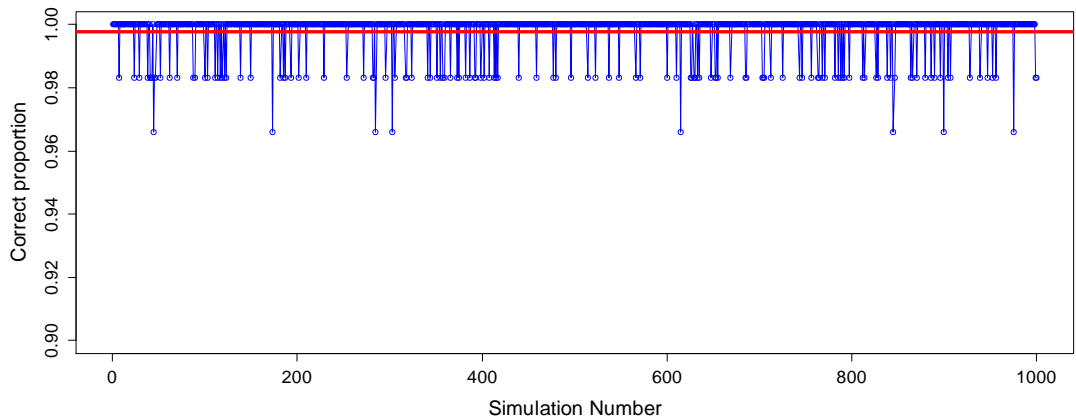

**(i) Correct re-link proportion in every simulation with SA1**

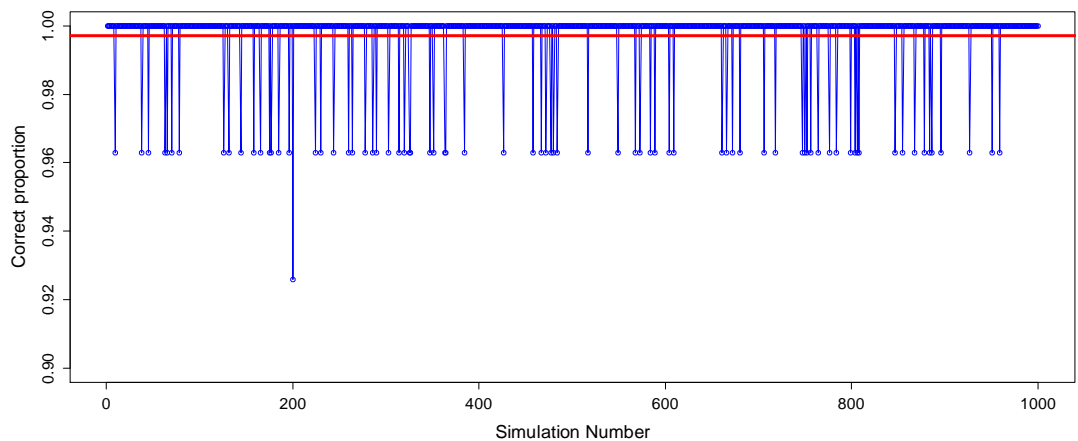

**(ii) Correct re-link proportion in every simulation with SA1 & SEX**

**Figure 4: Correct re-link proportion in every simulation**

In this analysis, the accuracy for all records in file *X* is estimated in every simulation. The plot (Figure 4 (i)) shows the correct re-link proportion of all 59 records in each of 1000 simulations when blocking with variable SA1. We obtained 100% accuracy in most of the simulations. In some simulations 98.3% accuracy is obtained where only one record is incorrectly re-linked out of 59 records. The smallest accuracy 96.6% (=57/59) is obtained in only eight simulations where two records are incorrectly linked. It is noticeable that the average accuracy in Figure 4 (i) for all records in every simulation is 99.75%, which is exactly the same as the average accuracy for each record in all simulations in Figure 3 (i).

When blocking with SA1 & SEX in Figure 4 (ii), we obtained 100% accuracy in most of the simulations. In some simulations 96.1% accuracy is obtained when 25 records (out of 26) are correctly linked to their original records. The smallest accuracy, 92.3% (=24/26), is obtained in only one simulation. Again, the average accuracy for all records in every simulation is 99.8%, which is the same as the average accuracy that is obtained previously for each record in all simulations in Figure 3 (ii).

When we block the data with SA1, with our original *MaCSim* we found the correct re-link proportion of each record in 1000 simulations lies between 93.5% and 100% whereas with extended *MaCSim*, it lies between 97% and 100%. Clearly, the assessment of accuracy of linkage improves. With combined variable SA1 & SEX, the correct re-link proportion of each record exceeds 98% using both approaches. We also examine the correct re-link proportion of all records in each of 1000 simulations using original and extended *MaCSim*. For single blocking variable SA1, with the original *MaCSim* the smallest accuracy, 93.2% (=55/59), is found in only three simulations where 4 records are incorrectly linked while with extended *MaCSim* the smallest accuracy 96.6% (=57/59) is obtained in only eight simulations where two records are incorrectly linked. For combined variable SA1 & SEX using extended *MaCSim*, the average accuracy for all records is 99.8%, which is an increase by 0.1% from using original *MaCSim*.

Other variables e.g., MB, BYEAR, BDAY can also be used for blocking and testing. When we blocked with these variables, we had only a small number of records to link (for MB, there were 6 records in one block) and the correct re-link proportions for those records exceeded 96%. Considering the purpose of the proposed method, we elected to present results with only a couple of blocking variables.

The extended *MaCSim* is implemented in the freely available statistical software R and the computational aspect of the methodology is investigated. The code is stable, parameterized and reusable on different sets of data. Blocking is used to reduce computational time. The computational time can be further reduced by using high performance computing (HPC) and applying optimization techniques, such as parallelisation. For a dataset, it is possible to save the overall simulated outcome and reuse it for analysis instead of going through the simulation process each time. The simulated agreement matrix can be stored and reuse to assess other linking methods.

## 5. Conclusion

To reduce the risk of missing potential matches from typographical or spelling errors, non-standardized formats of data etc., this paper introduced the use of partial agreement of the linking variable values in the form of a similarity weight in the extended *MaCSim*. Adding this feature to the original *MaCSim* is not expensive with respect to computation time but provides greater accuracy than obtained using the original algorithm.

The extended *MaCSim* is tested on numeric data for the case study. The method compares records (numeric values for the case study), checks for similarity and assigns values (from 0 to 1, -1 for missing) according to the match and nonmatch outcomes between records. *MaCSim* can also be used on text data fields, in which case data need to be prepared, such as by exploiting text similarity functions, the processes of parsing, standardisation etc.

The extended *MaCSim* measures the average accuracy of each record in all simulations and average accuracy for all records in every simulation. In both cases, more than 99% accuracy is obtained which indicates superior performance compared with the original *MaCSim*. Therefore, the extended *MaCSim* is proposed as a better method for assessing linking methods.